\documentstyle{aipproc}

\def\NPB{{\em Nucl. Phys.} B }
\def\PLB{{\em Phys. Lett.} B }
\def\PRL{{\em Phys. Rev. Lett.} } 
\def\PRD{{\em Phys. Rev.} D }
\def\PR{{\em Phys. Rev.} } 
\def\DKD{{\em Det. Kong. Danske Videnskabernes Selskab
Mat.-fysiske Meddelelser} } 
\def\PRS{{\em Proc. Roy. Soc. (London)} A }
\def\JETP{{\em JETP Lett.} }

\def\ga{\gamma}
\def\de{\delta}
\def\ep{\epsilon}

\def\la{\lambda}

\def\ph{\phi}

\def\ch{\chi}
\def\ps{\psi}

\def\Ga{\Gamma}
\def\De{\Delta}

\def\La{\Lambda}

\def\cl{{\cal L}}
\def\fr#1#2{{{#1} \over {#2}}}
\def\prt{\partial}

\def\vev#1{\langle {#1}\rangle}

\def\frac#1#2{{\textstyle{{#1}\over {#2}}}}

\def\lsim{\mathrel{\rlap{\lower4pt\hbox{\hskip1pt$\sim$}}
    \raise1pt\hbox{$<$}}}
\def\gsim{\mathrel{\rlap{\lower4pt\hbox{\hskip1pt$\sim$}}
    \raise1pt\hbox{$>$}}}
\def\sqr#1#2{{\vcenter{\vbox{\hrule height.#2pt
         \hbox{\vrule width.#2pt height#1pt \kern#1pt
         \vrule width.#2pt}
         \hrule height.#2pt}}}}

\def\Re{\hbox{Re}\,}
\def\Im{\hbox{Im}\,}

\newcommand{\beq}{\begin{equation}}
\newcommand{\eeq}{\end{equation}}
\newcommand{\bea}{\begin{eqnarray}}
\newcommand{\eea}{\end{eqnarray}}
\newcommand{\rf}[1]{(\ref{#1})}

\begin{document}

\begin{flushright}
{IUHET 370\\}
{August 1997\\}
\end{flushright}

\vglue 1cm

\title{Testing CPT with B Mesons}

\author{V. Alan Kosteleck\'y}
\address{Physics Department, Indiana University, Bloomington, IN 47405}

\maketitle

\begin{abstract}
Apparent violations of CPT and Lorentz symmetry
might arise in nature 
as a result of spontaneous symmetry breaking 
in a theory beyond the standard model.
This talk summarizes a few relevant 
theoretical and experimental issues,
with some emphasis on implications 
for CPT tests with neutral-$B$ mesons.
\end{abstract}

\section*{Introduction}

In this talk,
a brief survey is provided 
of theoretical and experimental issues relevant
to the notion that spontaneous CPT and Lorentz breaking 
might occur in a theory underlying the standard model.
The primary focus in the present work
is the possibility that observable effects 
could appear at accessible energies
from apparent CPT violation of this type.

For local relativistic field theories of point particles,
the CPT theorem
\cite{cpt1,cpt2,cpt3,cpt4,cpt5,cpt6,cpt7}
states that 
the combination of the three discrete operations of  
charge conjugation C, parity reflection P, and time reversal T
is an exact symmetry.
Various experiments have tested this theorem to high precision 
\cite{pdg}.
The existence of a powerful theorem 
and a wide variety of accurate experimental tests
suggests that apparent CPT violation
is a promising signal for new physics,
such as might emerge from a fundamental theory
beyond the standard model
\cite{kp1,kp2,kp3}.
At present,
string theory is the most promising approach to the construction 
of a consistent and complete quantum theory
of all fundamental particles and interactions.
However,
strings are extended objects,
so the conventional axioms underlying 
the proof of the CPT theorem for particle models 
do not necessarily apply.

In the next section,
a possible mechanism arising within string theory 
for spontaneous 
CPT \cite{kp1,kp2}
and 
Lorentz \cite{ks}
breaking is briefly summarized.
The mechanism can be studied explicitly
in certain theories
\cite{ks2,kp4}.
If spontaneous CPT breaking does arise in a realistic
fundamental model,
it might be apparent in nature
at presently accessible energies.
The subsequent section
outlines the results of an effective-theory approach
to incorporating possible CPT-violating interactions
in a low-energy model.
A general extension of the standard model 
\cite{kp3,cksm}
is described that includes extra terms 
originating from spontaneous CPT and Lorentz breaking
but maintains the usual gauge invariances 
and power-counting renormalizability.

Experimental implications of 
spontaneous CPT and Lorentz breaking
can be investigated 
within the framework of this extension of the standard model.
In particular,
quantitative constraints can be placed
on the occurrence of CPT and Lorentz breaking in nature
by bounding the coefficients of the extra terms in the theory.
The remaining parts of this talk summarize 
a few of these possible experimental tests of CPT.
Some of the most sensitive searches for apparent CPT breaking 
can be performed in neutral-meson systems.
The associated flavor oscillations provide interferometric 
tests of CPT symmetry
in the two $B$ systems \cite{kp3,ck1,kv}
and the $D$ system \cite{kp3,ck2}
as well as in the more conventional setting of 
the $K$ system \cite{kp1,kp2,kp3}.
Other sensitive measures of CPT breaking also exist.
They include comparative measurements 
of anomalous magnetic moments 
of electrons and positrons
or of protons and antiprotons
\cite{bkr}.
It is also possible that
CPT breaking is important for baryogenesis
\cite{bckp}.

\section*{Spontaneous CPT and Lorentz Violation}

The most natural mathematical setting for string theories
appears to involve more than four spacetime dimensions.
Assuming the fundamental theory underlying nature 
includes higher dimensions and is Lorentz and CPT symmetric,
then it is plausible that 
the higher-dimensional Lorentz invariance 
is spontaneously broken to the four observed dimensions.

A mechanism that could trigger this effect 
is known in string theory
\cite{ks}.
The nonlocality of strings generates interactions
in string field theory that do not appear in the context of
usual renormalizable four-dimensional gauge theory
but that are compatible with 
the infinite number of particle fields
and the string gauge invariances.
If appropriate scalar fields in the theory
acquire nonzero vacuum expectation values,
the static interaction potentials for Lorentz tensor fields
can be destabilized by stringy interactions of this kind.
Some of these Lorentz tensors may then obtain expectation values,
so that Lorentz invariance is spontaneously broken
in the true ground state of the theory.
It can be shown that 
CPT is also spontaneously broken
if any Lorentz tensor field 
with an odd number of spacetime indices 
acquires an expectation value 
\cite{kp1,kp2}.

The string field theory of the open bosonic string
provides an explicit example within which 
the mechanism for spontaneous Lorentz and CPT violation
can be studied.
A level-truncation scheme permits a systematic
exploration of the possible extrema of the action
\cite{ks2,kp4}.
It is feasible to construct the action analytically 
by incorporating only particle fields 
with level number less than a chosen value $N$.
The equations of motion can subsequently be found 
and solved for extrema of the action.
The procedure can be repeated 
and the solutions obtained can be compared
for different values of $N$.
A given solution consists of a definite set
of nonzero expectation values.
It is of interest if it persists 
and appears to converge as $N$ increases,
since it is then plausible that 
the complete theory contains an extremum 
involving similar expectation values.

Following this procedure 
and using symbolic manipulation routines,
it has been feasible
to study aspects of the static interaction potential
for the open bosonic string
to a depth of over 20,000 nonzero terms.
Nontrivial solutions to the equations of motion emerge,
including ones violating Lorentz and CPT invariance.
These exhibit properties that are to be expected
from general considerations of the theoretical mechanism.

\section*{Extension of the Standard Model}

A question of immediate relevance is whether 
the occurrence of spontaneous CPT and Lorentz breaking
in a fundamental theory could generate
apparent violations at low energies.
This would happen if the breaking extends
to the four large spacetime dimensions,
which is natural mathematically.
Otherwise,
a definite mechanism would be needed 
to explain the existence of the four-dimensional symmetry.

No CPT or Lorentz violations have been found experimentally.
A high degree of suppression is therefore implied
for any possible effects at low energies.
The standard model is known to provide an excellent
description of nongravitational physics in this regime.
It can be viewed as an effective model 
emerging at the electroweak scale $m_{\rm ew}$
from a more fundamental theory,
presumably governed by the Planck scale $m_{\rm Pl}$.
The natural suppression factor
for Planck-scale effects in the standard model 
would then be
$r \sim m_{\rm ew}/m_{\rm Pl} \simeq 10^{-17}$.
Relatively few CPT- and Lorentz-violating effects 
would be accessible to experiment 
if this strong suppression occurs. 
The sections below describe a few of the possible signals.

Suppose the spontaneous CPT and Lorentz violation 
indeed generates minuscule effects at the level of 
the electroweak scale.
These effects can be studied by extending the standard model
to include possible terms 
originating in spontaneous CPT and Lorentz violation.
In the fermion sector,
for example,
possible terms of the general form \cite{kp2,kp3}
\beq
\cl \sim \fr {\la} {M^k} 
\vev{T}\cdot\overline{\ps}\Ga(i\prt )^k\ch
+ {\textstyle h.c.}
\quad 
\label{a}
\eeq
could appear as a low-energy consequence
of spontaneous symmetry breaking in a compactified string theory.
Terms of this type are Lorentz- and possibly CPT-violating,
as a result of the nonzero expectation values of 
Lorentz tensors $T$ that can appear in interactions terms
coupling $T$ with fermions $\ps$ and $\ch$ 
via derivatives $i\prt$ and a gamma-matrix structure $\Ga$.
In Eq.\ \rf{a},
$\la$ is assumed dimensionless,
so one or more large mass scales $M$ 
such as the compatification or Planck scales
must also appear.
The fermions $\ps$ and $\ch$ in Eq.\ \rf{a}
can be identified with leptons or quarks
in the standard model.

A general extension of the standard model
has been obtained
\cite{cksm}
that includes Lorentz-breaking terms
both with and without CPT breaking
and allows also for possible effects in the gauge and Higgs sectors.
The analysis is developed around a theoretical framework
for treating CPT and Lorentz breaking
that appears to bypass some standard difficulties,
largely because the breaking is spontaneous
while the underlying theory remains Lorentz and CPT invariant.
The derivation therefore includes the constraint 
that any new effects must be compatible 
with an origin in spontaneous Lorentz breaking.
Invariance under the gauge group
SU(3) $\times$ SU(2) $\times$ U(1)
and power-counting renormalizability are also required.

The next sections
outline some possible measurable signals
that are implied by this standard-model extension.
Detailed treatments exist at present
for effects in neutral-meson systems
and for effects in certain experiments
in the context of quantum electrodynamics.

\section*{Tests with Neutral Mesons}

In the four neutral-meson systems,
$K$, $D$, $B_d$, $B_s$,
the small mass differences 
between weak-interaction eigenstates 
offer an interferometric sensitivity
to highly suppressed effects
such as might arise from Planck-scale physics
\cite{kp3}.
This section has three subsections.
One summarizes theoretical issues,
one outlines general experimental issues 
and some established results, 
and the third discusses the case of the $B_d$ system. 

\subsection*{Theoretical Issues}

The time evolution of a neutral-$P$ meson,
where the symbol $P$ denotes any of the four neutral mesons,
is governed by a $2\times 2$ effective hamiltonian $\La_P$.
Assuming conventional quantum mechanics,
there are two types of indirect CP violation
that might be described by 
phenomenological parameters appearing in $\La_P$.
The first is the usual CP-violating parameter $\ep_P$
that breaks T but preserves CPT.
The second is a complex CP-violating parameter $\de_P$
that preserves T but breaks CPT.
Note that this parametrization of CP violation 
is independent of any underlying model
and is at a purely phenomenological level.

In the context of the usual standard model,
parameters in the CKM matrix can be regarded as 
controlling the parameters $\ep_P$ for each $P$.
No such understanding exists 
for possible nonzero values of $\de_P$.
However, 
the CPT-violating extension of the 
standard model outlined in the previous section 
does provide a theoretical basis for nonzero $\de_P$.
For instance,
terms of the type shown in Eq.\ \rf{a},
which appear in the quark sector 
of the general standard-model extension
when the fermions $\ps$ and $\ch$ are identified with quark fields,
act to change the time evolution of a neutral-$P$ meson
in a $\de_P$-dependent way.
The point is that the propagators of the (valence) quarks 
are affected and generate CPT-violating effects.

A general expression for the
the quantity $\de_P$ for a given $P$ system
can be obtained
in the context of spontaneous CPT and Lorentz breaking.
It is \cite{kp2,kp3}:
\beq
\de_P = i 
\fr{h_{q_1} - h_{q_2}}
{\sqrt{\De m^2 + \De \ga ^2/4}}
e^{i\hat\ph}
\quad .
\label{b}
\eeq
In this equation,
$\De m$ and $\De\ga$ are the $P$-meson mass and rate differences,
which are experimentally observable.
They are used to define
the angle $\hat\ph$ by $\hat\ph \equiv \tan^{-1}(2\De m/\De \ga)$.
The quantities $h_{q_j}$ 
are determined by parameters in the standard-model extension 
arising from the spontaneous CPT violation in the fundamental theory
and by the effects $r_{q_j}$ of the quark-gluon sea:
$h_{q_j}=r_{q_j}\la_{q_j}\vev{T}$.

The hermiticity of the underlying theory
and the extension of the standard model
implies that the quantities $h_{q_j}$ are real.
This in turn can be used to show that 
the real and imaginary parts of $\de_P$ are proportional,
with proportionality constant determined
by experimentally measurable rate and mass differences.
Explicitly,
the result is
\beq
\Im \de_P = \pm \fr{\De\ga}{2\De m} ~\Re\de_P
\quad . 
\label{c}
\eeq
This can be regarded as a determining signature
for CPT breaking arising within the present framework.
Assuming a suppression factor of $r\simeq 10^{-17}$
along the lines of the discussion above,
it also follows that direct CPT violation
arising in the $P$-meson decay amplitudes is 
too small to observe.

Within the present framework,
it is plausible that
the values of the CPT-violating quantities $\de_P$ 
could be significantly different 
for distinct $P$-meson systems.
The point is that the dimensionless coupling constants 
$\la_{q_j}$,
appearing in terms of the form given in Eq.\ \rf{a},
might depend on the quark flavor $q_j$.
The corresponding CPT-violating quark couplings 
within the extension of the standard model
would then also be flavor dependent.
A related effect occurs for the Yukawa couplings,
which take values 
for different quark flavors 
that range over about six orders of magnitude.
The possibility of flavor-dependent CPT violation 
means that the values of $\de_P$ might vary with $P$,
so it might be crucial to perform
experimental tests of CPT symmetry
in more than one neutral-meson system.
Moreover,
under some circumstances the experimental signals 
could be startling.
For instance,
only relatively weak limits have been obtained as yet
on $B_d$-meson CP violation,
which means it is possible that
CP violation parametrized by the CPT-violating quantity $\de_{B_d}$
could be \it larger \rm than
conventional CP breaking parametrized by $\ep_{B_d}$
and therefore could produce unexpected results
in proposed experiments at $B$ factories.

\subsection*{Experimental Issues and Results}

This subsection provides a short outline 
of some experimental issues and the present status
of established tests of indirect CPT violation with
neutral-$P$ mesons. 
Tests with neutral-$B$ mesons
are discussed in the following subsection. 

Searches for indirect CP violation in a neutral-$P$ system
can be performed with experimental data taken
from decays either of uncorrelated tagged $P$ mesons
or of correlated $P$-$\overline P$ pairs 
produced through prior quarkonia decay.
Both indirect T and CPT violation
are in principle accessible.
Time-dependent and fully integrated
decay-probability asymmetries,
sensitive to the various CP parameters,
have now been established in each $P$-meson system
for both correlated and uncorrelated situations. 
These can be applied to analyses of real data,
or can be used in theoretical estimates of CP reach
performed either analytically or 
through detailed Monte-Carlo simulations
with acceptances and background effects.

To date,
the sharpest bound on CPT violation 
in a neutral-meson system
comes from analyses of $K$ oscillations.
There already exist published bounds
on $|\de_K|$ of order $10^{-3}$
\cite{pdg,expt1,expt2}.
Analyses currently underway 
or experiments being performed or planned 
are anticipated to provide even tighter limits.

In the $D$ system,
mass mixing has yet to be detected.
Furthermore,
the expected strong dispersive effects
and the complication of dominant contributions 
arising from physics beyond the standard model
makes theoretical predictions difficult
and subject to uncertainties potentially of
orders of magnitude.
Nonetheless,
in circumstances that are favorable theoretically,
certain tests of CPT invariance
in the $D$ system might produce signals 
when performed with current techniques 
and perhaps even with data that already exist
\cite{ck2}.
The expected increase in reconstructed events
to be obtained in various future machines
provides an interesting arena for 
establishing CPT bounds from the $D$ system.

\subsection*{The Neutral-$B$ System}

Tests of CPT with neutral-$B_d$ mesons are 
of interest both theoretically and experimentally.
Theoretically,
if indeed the dimensionless couplings 
corresponding to $\la$ in Eq.\ \rf{a}
are flavor dependent 
and follow the same general pattern as the 
Yukawa couplings,
then the strength of the CPT violation
is related to the mass of the valence quarks 
that are bound in the neutral meson.
In this case,
since the $b$ quark is involved,
it is possible that any CPT violation would be larger
in the $B$ system than in other neutral-meson systems
\cite{kp3,kv}.
On the experimental front,
large numbers of $B_d$ events have already been obtained,
and future machines and detectors under construction
are expected to produce high-statistics event samples.

Several studies have been performed 
to estimate the likely constraints 
on CPT violation that could be obtained from
experiments with $B_d$ mesons 
\cite{kp3,ck1,kv}.
The most detailed treatment uses Monte-Carlo simulations
to model experiments performed
with uncorrelated, correlated unboosted,
or correlated boosted mesons
\cite{kv}.
Backgrounds, resolutions, and acceptances 
are incorporated in simulating realistic experimental data
that might be obtained at typical detectors 
at LEP, CESR, and the future $B$ factories.
One result from this analysis is 
that data already taken suffice to place meaningful bounds 
on $\de_{B_d}$.

Until recently,
no bound existed on $\de_{B_d}$.
Early in 1997,
the OPAL collaboration at LEP
obtained the first experimental constraint
on CPT violation in the neutral-$B_d$ system
\cite{expt3}.
The relevant experimental observable 
is an asymmetry derived in Ref.\ \cite{kv}
that is sensitive to 
$\Im\de_{B_d}$ and $\Re\ep_{B_d}$.
The time evolution of this asymmetry can be 
extracted from the experiment and 
used to bound both these quantities. 
The result is a constraint on the value of $\Im\de_{B_d}$ 
of less than $3 \times 10^{-2}$
at the 95\% confidence limit.

An interesting feature of the $B_d$ system
is that Eq.\ \rf{c}
predicts that the real part of $\de_{B_d}$
is greater than the imaginary part
because the value of $2\De m/\De\ga$ 
is believed to be large. 
In contrast,
the real and imaginary parts of $\de_P$ for the
$K$ system and perhaps also the $D$ system would be comparable. 
The analysis of Ref.\ \cite{kv}
suggests that data already taken with the CLEO detector at CESR
could be used to bound $\Re\de_{B_d}$.
The expected relatively large size of this quantity
compared to $\Im\de_{B_d}$
implies that even a limit of order 20\% 
would be of interest.

The $B$ factories and other $B$-dedicated experiments
now under construction 
should be capable of improving on bounds  
obtained from current data.
Moreover,
the corresponding detectors are also expected
to be sensitive to both 
$\Re\de_{B_d}$ and $\Im\de_{B_d}$
\cite{kv}.
This opens the possibility in principle 
of testing Eq.\ \rf{c} in a single experiment,
should CPT violation indeed be discovered.

\section*{Effects in Other Systems}

Effects from spontaneous CPT and Lorentz violation
could also be manifest 
in contexts other than neutral-meson oscillations.
In the standard-model extension,
distinct quantities govern CPT and Lorentz breaking in,
for example,
the quark and lepton sectors.
A wide variety of experiments outside the neutral-meson systems
is therefore potentially crucial for uncovering effects.
This section briefly describes 
some possibilities of this type.

The standard description of baryogenesis
requires CP- and C-violating interactions
and nonequilibrium processes 
\cite{ads}
as well as baryon-number violating effects.
In grand-unified theories, 
for example,
the CP breaking is selected in a range suitable
for reproducing the known baryon asymmetry
and is unrelated to the observed CP violation
in the standard model.
The presence of CPT-breaking processes
of the type given in Eq.\ \rf{a}
suggests an alternative possibility for baryogenesis
that could occur in thermal equilibrium 
without the need for additional CP violation.
An analysis \cite{bckp} shows that
under suitable circumstances
a mechanism of this kind
could result in a large baryon asymmetry
at grand-unification scales
that diminishes to the observed value
by a process such as sphaleron dilution.

Another implication of the CPT- and Lorentz-violating
extension of the standard model
is a modification of some conventional results in 
quantum electrodynamics
\cite{cksm}.
One example concerns comparative measurements
of the anomalous magnetic moments of the electron and positron.
It has recently been shown \cite{bkr}
that the standard figure of merit 
used in these experiments is misleading.
However, 
a more appropriate measure can be defined 
that is directly sensitive to 
some of the additional terms appearing  
in the modified version of quantum electrodynamics.
With current experimental techniques,
constraints on CPT violation could be attained
with a precision similar to those from neutral-meson systems.
Related experiments with protons and antiprotons
may provide interesting limits on CPT violation.
Bounds on CPT and Lorentz breaking are also possible 
from precision experiments 
using cyclotron frequencies
\cite{bkr}
and photon properties
\cite{cksm}.

\section*{Acknowledgments} 

I thank Orfeu Bertolami, Robert Bluhm, Don Colladay, 
Rob Potting, Neil Russell, Stuart Samuel, 
and Rick Van Kooten for collaborations.
This work was supported in part
by the United States Department of Energy 
under grant number DE-FG02-91ER40661.


\begin{references}

\bibitem{cpt1}
J. Schwinger, \PR {\bf 82} (1951) 914.

\bibitem{cpt2}
G. L\"uders, 
\DKD
{\bf 28}, no.\ 5 (1954).

\bibitem{cpt3}
J.S. Bell,
Ph.D.\ thesis (Birmingham University, England, 1954);
\PRS
{\bf 231} (1955) 479.

\bibitem{cpt4}
W. Pauli, in 
\it Niels Bohr and the Development of Physics, \rm
ed.\ W. Pauli,
(McGraw-Hill, New York, 1955), p.\ 30.

\bibitem{cpt5}
G. L\"uders and B. Zumino,
\PR 
{\bf 106} (1957) 385.

\bibitem{cpt6}
R.F. Streater and A.S. Wightman,
\it PCT, Spin and Statistics, and All That \rm
(Benjamin Cummings, Reading, 1964).

\bibitem{cpt7}
R. Jost,
{\it The General Theory of Quantized Fields}
(AMS, Providence, 1965).

\bibitem{pdg}
See, for example, R.M. Barnett {\it et al.},
{\it Review of Particle Properties,}
\PRD 
{\bf 54} (1996) 1.

\bibitem{kp1}
V.A. Kosteleck\'y and R. Potting,
\NPB 
{\bf 359} (1991) 545.

\bibitem{kp2}
V.A. Kosteleck\'y, R. Potting, and S. Samuel,
in 
\it Proceedings of the 1991 Joint International Lepton-Photon
Symposium and Europhysics Conference on High Energy Physics, \rm
eds.\ S. Hegarty {\it et al.}
(World Scientific, Singapore, 1992); 
V.A. Kosteleck\'y and R. Potting,
{\it Gamma Ray--Neutrino Cosmology and Planck Scale Physics,} \rm
ed.\ D.B. Cline
(World Scientific, Singapore, 1993)
(hep-th/9211116).

\bibitem{kp3}
V.A. Kosteleck\'y and R. Potting,
\PRD 
{\bf 51} (1995) 3923.

\bibitem{ks}
V.A. Kosteleck\'y and S. Samuel,
\PRD 
{\bf 39} (1989) 683;
\it ibid., \rm
{\bf 40} (1989) 1886;
\PRL {\bf 63} (1989) 224;
\it ibid., \rm
{\bf 66} (1991) 1811.

\bibitem{ks2}
V.A. Kosteleck\'y and S. Samuel,
\NPB 
{\bf 336} (1990) 263;
\PRL 
{\bf 64} (1990) 2238;
\PRD 
{\bf 42} (1990) 1289.

\bibitem{kp4}
V.A. Kosteleck\'y and R. Potting,
\PLB 
{\bf 381} (1996) 389.

\bibitem{cksm}
D. Colladay and V.A. Kosteleck\'y,
\PRD 
{\bf 55} (1997) 6760;
preprint IUHET 359 (1997),
\it Phys.\ Rev.\ D, \rm in press (hep-ph/9809521).

\bibitem{ck1}
D. Colladay and V.A. Kosteleck\'y,
\PLB 
{\bf 344} (1995) 259.
 
\bibitem{kv}
V.A. Kosteleck\'y and R. Van Kooten,
\PRD 
{\bf 54} (1996) 5585.

\bibitem{ck2}
D. Colladay and V.A. Kosteleck\'y,
\PRD 
{\bf 52} (1995) 6224.

\bibitem{bkr}
R. Bluhm, V.A. Kosteleck\'y, and N. Russell,
\PRL
{\bf 79} (1997) 1432;
\PRD 
{\bf 57} (1998) 3932.

\bibitem{bckp}
O. Bertolami 
{\it et al.}, 
\PLB 
{\bf 395} (1997) 178.

\bibitem{expt1}
L.K. Gibbons 
{\it et al.}, 
\PRD 
{\bf 55} (1997) 6625.

\bibitem{expt2}
R. Carosi 
{\it et al.}, 
\PLB 
{\bf 237} (1990) 303.

\bibitem{expt3}
OPAL Collaboration, R. Ackerstaff 
{\it et al.}, 
\it Z.\ Phys. C \rm {\bf 76} (1997) 401;
DELPHI Collaboration,
M.\ Feindt
{\it et al.},
preprint DELPHI 97-98 CONF 80 (July 1997).

\bibitem{ads}
A.D. Sakharov,
\JETP 
{\bf 5} (1967) 24.

\end{references}
\end{document}